\documentclass[epjST,final]{svjour}
\usepackage{graphics}
\usepackage{subfig}
\usepackage{url}

\newcommand{\bhu}{ \hat{\bf u} }

\newcommand{\br}{ {\bf r} }

\newcommand{\ga}{ {\alpha }}
\newcommand{\gb}{ {\beta }}

\begin{document}
\title{Motion of two micro-wedges in a turbulent bacterial bath }
\author{Andreas~Kaiser\inst{1} \and Andrey~Sokolov\inst{2} \and Igor~S.~Aranson\inst{2} \and Hartmut~L\"owen\inst{1}}
\institute{
Institut f\"ur Theoretische Physik II: Weiche Materie, 
Heinrich-Heine-Universit\"at D\"usseldorf, 
Universit\"atsstra{\ss}e 1,
40225 D\"usseldorf, Germany
\and
Materials Science Division, 
Argonne National Laboratory,
9700 South Cass Avenue, 
Argonne, Illinois 60439, USA
}

\abstract{ 
The motion of a pair of micro-wedges ("carriers") in a turbulent bacterial bath
is explored using computer simulations with explicit modeling of the bacteria and experiments. 
The orientation of the two micro-wedges is fixed by an external magnetic field but the translational 
coordinates can move freely as induced by the bacterial bath. As a result,
two carriers of same orientation move such that their mutual distance decreases, while they drift apart for an anti-parallel orientation.
Eventually the  two carriers stack on each other with no intervening bacteria
exhibiting a stable dynamical mode 
where the two micro-wedges follow each other with the same velocity. These findings are in qualitative agreement 
with experiment on two micro-wedges in a bacterial bath. 
Our results provide insight into understanding   self-assembly of many micro-wedges in an active bath.
} 
\maketitle
\section{Introduction}
\label{intro}

A wide variety of active suspensions~\cite{Romanczuk2012,aranson_ufn,Marchetti_Rev,Cates_2012} are known to form remarkable spatio-temporal 
patterns~\cite{Toner,Vicsek_Report2012,2011KochSub}
with the appearance of coherent dynamics structures on scales that are large compared with a single self-propelled unit.
Examples range from bacterial suspensions~\cite{drescher2011fluid,2007SoEtAl}, spermatozoa~\cite{2005Riedel_Science,Friedrich}, 
human crowds~\cite{Silverberg:13} to suspensions composed out of artificial self-propelled particles~\cite{kudrolli,Aranson_RMP,SenPNAS13,Sano_PRL2010,Baraban_ACSnano,KapralJCP2013}.
Such systems have frequently been studied in the last year in bulk focusing on clustering~\cite{2006Peruani,BocquetPRL12,PerunaiPRL12,Bialke_PRL2013,Palacci_science,ZoettelPRL14}, 
swarming~\cite{2009Swinney,2011Herminghaus,vicsek_prl,gompper2013} and complex swirling or turbulence~\cite{PNAS,2008Wolgemuth,ISA-PNAS14,DunkelPRL13,CollectiveSpheres,GrossmannTurbulence,Clement_NJP14,Sokolov_PRL12}.
A static confinement has been shown to be able to stabilize these structures~\cite{GoldsteinPNAS2014}, 
accumulate and guide active particles~\cite{2010Gompper,Wensink2008,li2009accumulation,DenissenkoPNAS,HaganConfinement,PoonRingConfinement}.
This effect has been used to rectify the motion of swimmers~\cite{Chaikin2007,Reichhardt_PRL2008,RectifyBrazil,StarkPRE13Rectification,CatesTrapping} 
and to build sorting~\cite{Reichhardt,hulme,MarconiSorting,JulichSortingMicrochannel}
as well as trapping devices~\cite{Kaiser_PRL,TrappingSperms,TrappingOGS}. Furthermore the motion of passive but mobile particles submersed in an active
fluid has been studied, starting with spherical and curved tracers~\cite{TracerGoldstein,TracerClement,MalloryArxiv} to long deformable chains~\cite{KaiserPolymer}.
Using asymmetric cogwheels a spontaneous directed rotation~\cite{SokolovPNAS,LeonardoPNAS,GearRobots} 
can be extracted out of active bath. The translational analog is a directed motion of a single micro-wedge along its cusp
induced by the active particles~\cite{KaiserSokolov_2014,AngelaniCargo}.

In this paper, we consider micro-wedges as passive carriers and expose them to a turbulent bacterial bath. 
The case of a single carrier has been explored previously both by computer simulation of appropriate models 
resolving the individual bacteria and by experiments~\cite{KaiserSokolov_2014}.
For a micro-wedge kept fixed in orientation by an external uniform field and moving on a two-dimensional substrate as far 
as its translational motion is concerned, it was found that turbulence of the bath
maximizes the directed carrier speed along its cusp. The responding mechanism was
attributed to swirl depletion in the inner wedge area which gives some
bacteria which are close to the wedge angle the possibility to push the
carrier forward in an efficient way.

Here, we consider the case of two micro-wedges with the fixed orientation
and explore by computer simulation and experiments the
 motion of a pair of carriers in two dimensions.
This can be understood as a first step towards the
hierarchical self-assembly of many carriers in a bacterial bath. 
We compare two different configurations of the wedges: parallel and anti-parallel orientated
carriers. Since each wedge is transported in the direction of its cusp, two carriers of the same orientation move such that their mutual distance decreases, 
while they drift apart for an anti-parallel orientation.
Eventually, the two carriers of same orientation will end up
in a state where they stack closely on top of each other such that there are no intervening
bacteria left. 
They exhibit a stable dynamical mode in which the two
micro-wedges follow each other with the same speed.
The distribution function of the wedge distance averaged over a finite time shows a subtle multiple-peak structure
which is compatible with the swirl depletion picture.
We obtain our results within the same model successfully
applied to the description of a single micro-wedge by calculating the average
relative velocity and the distribution of the wedge distances. Our numerical results are in qualitative agreement with experiment on 
two micro-wedges in a turbulent bacterial bath.

This paper is organized as follows: in Sect.~\ref{model} we will explain the used model before we present results
for the motion of the micro-carriers obtained by particle resolved simulations in Sect.~\ref{resultsSim} and experiments in Sect.~\ref{resultsExp}. 
Finally, we conclude in Sect.~\ref{conclusions}.

\section{Model}
\label{model}

\begin{figure}
\centering
\resizebox{0.8\columnwidth}{!}{\includegraphics{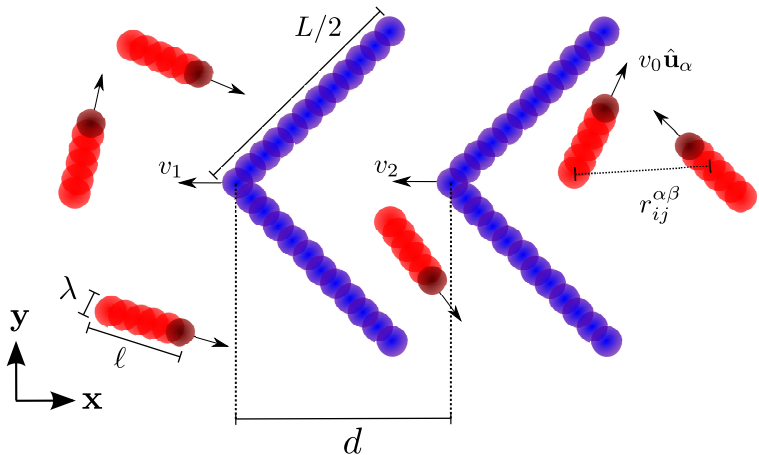}}
\caption{Schematic representation of the system. The self-propelled rods with an aspect ratio $\ell / \lambda$ have a propulsion velocity $v_0$ directed along the main rod axis $\bhu$. The single six Yukawa segments are shown by red circles, whereby the larger interaction prefactor of the first segment is indicated by darker color. The Yukawa segments of the
wedgelike carriers are shown in blue. The distance between the carriers is given by $d$ and $L$ corresponds to the contour length of each carrier. On average their velocity will be directed along the cusp and are denoted by $v_1$ and $v_2$.}
\label{fig:Sketch}   
\end{figure}

We model the bacterial bath in two spatial dimensions by $N$ rod-like self-propelled units with
an effective body shape asymmetry analogous to Ref.~\cite{KaiserSokolov_2014}. Each rod
of length $\ell$ and width $\lambda$ is discretized into $n=6$ spherical segments equidistantly positioned, with a displacement $s=0.85\lambda$,
along the main rod axis $\bhu = (\cos \varphi, \sin \varphi)$, see Fig.~\ref{fig:Sketch}. The according aspect ratio $p= \ell / \lambda = 5$ is chosen
in order to model {\it Bacillus subtillis} suspensions, as considered in our experiments. A repulsive Yukawa potential is imposed 
between the segments of different rods~\cite{Kirchhoff1996}. The resulting pair potential of a rod pair $\alpha$, $\beta$ is given by
$U_{\alpha \beta} = \sum_{i=1}^{n} \sum_{j=1}^{n} U_i U_j \exp [-r_{ij}^{\alpha \beta} / \lambda] /  r_{ij}^{\alpha \beta} $, where $\lambda$ is the
screening length and $ r_{ij}^{\alpha \beta} = |{\bf r}_{i}^{\ga} - {\bf r}_{j}^{\gb}|$ the distance between segment
$i$ of rod $\alpha$ and segment $j$ of rod $\beta$ $(\alpha \neq \beta)$.
In analogy to our previous work~\cite{KaiserSokolov_2014,KaiserIEEE} we incorporate an effective shape asymmetry to account for the experimental observed
swim-off effect of colliding bacteria~\cite{drescher2011fluid,2007SoEtAl,aranson2007}. We increase the interaction prefactor 
for the first segment with respect to the others of each rod. This quantity is given by $U_1^2/U_j^2 = 3$ $(j = 2\ldots n)$~\cite{KaiserSokolov_2014}. 
Any overlap of particles is prohibited by choosing a large interaction strength $U_j^2=2.5F_0 \ell$. Here $F_0$ is an effective self-propulsion
force directed along the main rod axis and leading to a constant propulsion velocity $v_0$~\cite{BtHComment}. We do not resolve  details
of the actual propulsion mechanism. 
Hydrodynamic interactions between the swimmers are neglected which is expected to be justified at high 
packing fractions in the absence of any global flow, i.e. in an orientationally disordered configuration as considered here~\cite{GoldsteinPNAS2014}.

Micro-swimmers move in the low Reynolds number regime. The corresponding overdamped equations of motion for the positions 
$\br_{\alpha}$ and orientations $\bhu_\alpha$ are

\begin{eqnarray}
{\bf f }_{\cal T} \cdot \partial_{t} \br_{\alpha}(t) &=&  -\nabla_{\br_{\alpha}}
 U(t) +  F_{0} \bhu_{\alpha}(t), \\
{\bf f}_{\cal{R}} \cdot \partial_{t} \bhu_{\alpha}(t) &=&
-\nabla_{\bhu_{\alpha}} U(t),
\label{eom}
\end{eqnarray}

in terms of the total potential energy  
$U=(1/2)\sum_{\alpha, \beta (\alpha \neq \beta)} U_{\alpha \beta} + \sum_{\alpha, \gamma} U_{\alpha \gamma}$ 
with $U_{\alpha \gamma}$ the potential energy of rod $\alpha$ with the carrier $\gamma$. 
The one-body translational and rotational friction tensors for the rods ${\bf f}_{\cal T}$ 
and ${\bf f}_{\cal R}$ can be decomposed into parallel $f_\parallel$, perpendicular $f_\perp$
and rotational $f_{\cal R}$ contributions which depend solely on the aspect ratio $p = \ell/\lambda $~\cite{tirado,HL_PRE_1994}

\begin{eqnarray}
\frac{2\pi}{f_{||}} &=& \ln p - 0.207 + 0.980p^{-1} - 0.133p^{-2},
\\
\frac{4\pi}{f_{\perp}}&=&\ln p+0.839 + 0.185p^{-1} + 0.233p^{-2},
\\
\frac{\pi a^2}{3f_{\mathcal{R}}} &=& \ln p - 0.662 + 0.917p^{-1} - 0.050p^{-2}.
\end{eqnarray} 

Accordingly, the propulsion velocity is given by $v_0 = F_0 / f_{||}$ and sets the characteristic
time unit $\tau = \ell / v_0$.

We model a pair of micro-wedges $\gamma$, $\delta$ analogous to the swimmers by tiling the wedge contour length $L=20\ell$
into Yukawa segments and restrict their motion to translation. The wedge angle is kept rectangular.
The resulting equation of motion for the carriers is

\begin{eqnarray}
 {\bf f}_{\gamma} \cdot \partial_t {\bf r}_{\gamma}(t) = - \nabla_{{\bf r}_{\gamma}} U_{\gamma}(t),
\end{eqnarray}
where ${\bf f}_{\gamma}$ corresponds to the hydrodynamic friction coefficient of the micro-wedge, calculated using the software package \texttt{HYDRO++}~\cite{delaTorreNMDC1994,Carrasco99} 
and $U_\gamma =  (1/2)\sum_{\gamma, \delta (\gamma \neq \delta)} U_{\gamma \delta} + \sum_{\gamma, \alpha} U_{\gamma \alpha}$ is the total interaction on a single micro-wedge.

We use a rectangular simulation domain with $L_y = 3L / \sqrt{2}$, an aspect ratio $L_x / L_y = 3$ and an area $A=L_x L_y$ and periodic boundary conditions.
The total number of rods is given by $N = A\phi / \lambda \ell$, where $\phi$ is the dimensionless packing fraction, which will be fixed to $\phi=0.5$
to achieve a turbulent bacterial bath~\cite{PNAS,KaiserSokolov_2014,WensinkJPCM}. 
Results are obtained for simulations with randomly chosen starting distances $d$ and statistics are being gathered over a time interval $t = 1000\tau$, using a time step $\Delta t = 10^ {-3} \tau$. By measuring the mean displacement along the {\bf x}-direction within a single time step, the individual velocities $v_\gamma$ are determined by
$v_{\gamma} = \left(  x_\gamma (t+\Delta t) - x_\gamma (t)  \right) / \Delta t $, with $\gamma = 1,2$.
The obtained results and the resulting transport efficiencies for the carriers  depend weakly on the packing fraction within the turbulent regime of the bacterial bath~\cite{KaiserIEEE}. 

\section{Simulation results}
\label{resultsSim}
\begin{figure}
\centering
\resizebox{1.0\columnwidth}{!}{\includegraphics{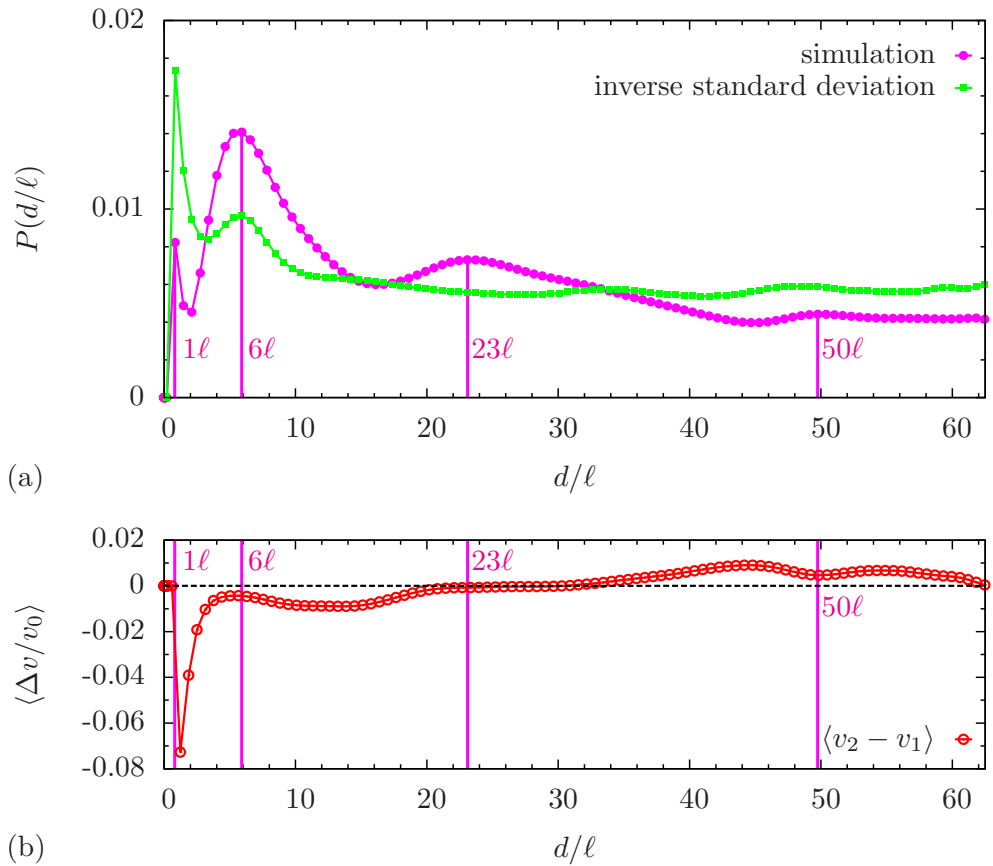}}
\caption{(a) Probability distribution of the carrier distance $d$ measured in swimmer lengths $\ell$ obtained from simulations (magenta) 
and predictions due to the achieved transport efficiencies of the carriers. (b) Difference of the transport efficiencies $\Delta v=v_2-v_1$ 
of the single carriers. Negative values mean that the right carrier is catching up, see Fig.~\ref{fig:Sketch}. Vertical lines indicate the 
location of the local maxima in the probability distribution $P(d/\ell)$.}
\label{Px}   
\end{figure} 

In the following we consider two cases. Firstly,  we confine the motion of the two wedges under the constraint that
they both have the same $y$-coordinate and their orientation is fixed while their 
$x$-coordinates can respond freely to the bacterial bath. 
Secondly,  we only fixed their orientation but relax the constraint in the $y$-direction. In our experiment the alignment constraints were 
imposed by the external uniform magnetic field.

For the first case, Fig.~\ref{Px}(a) shows the probability distribution $P(d/\ell)$ for the distance between two micro-wedges 
of same orientation, using the reduced distance $d/\ell$ between the carriers, measured along the cusp of the carriers, 
see again Fig.~\ref{fig:Sketch}. The distribution reveals four characteristic peaks at distances $d_1 = 1\ell$, $d_2 = 6\ell$,
 $d_3 = 23\ell$ and $d_4 = 50\ell$ and will be explained step by step in the following.

It is interesting to correlate the peak positions with the behavior of the 
transport speed difference  $\Delta v=v_2 - v_1$ as a function of the carrier distance $d$,  see
Fig.~\ref{Px}(b). In the absence of  velocity fluctuations, a
  peak in $P(d/\ell)$ is expected either when the transport speed difference vanishes 
or when the modulus of the speed difference
exhibits a local minimum. In case of a vanishing relative speed at $d=d_0$, the sign of the slope 
$\partial \Delta v/ \partial d|_{d=d_0}$ determines the stability of the stationary situation at $d=d_0$: it
 is stable if  $\partial \Delta v/ \partial d|_{d=d_0} <0$   and  unstable if $\partial \Delta v/ \partial d|_{d=d_0}>0$.
For a stable situation and in the absence of velocity fluctuations, the particle would be stuck 
at the distances where the velocity is vanishing
resulting in a divergence of the distribution function  $P(d/\ell)$ at these distances.

In Fig.~\ref{Px}(b) we observe indeed two zeroes at about $d_1 = 1\ell$ and about $d = 28\ell$ which 
compare with the peaks  at  $d_1 = 1\ell$ and $d_3 = 23\ell$. 
Moreover two minima in the speed difference occur at  $d_2 = 6\ell$ and  $d_4 =50\ell$ which clearly correlates
with the second and fourth peak in $P(d/\ell)$. However, for the actual height of the peak
velocity fluctuations are significant which smear out the "ideal" divergence. These are defined as
$\sqrt{\langle (\Delta v/v_0)^2 \rangle}$, where $\langle \ldots \rangle$ denotes a time average, 
and shown in Fig.~\ref{Px}(a) as well and reveal a non-Brownian behavior.

In detail, the first peak at $d_1 = 1\ell$
where the two wedges stick together corresponds indeed 
to a stable situation.
When compressing the wedges more to an even smaller distance than $d_1 = 1\ell$,
 the repulsive bead forces acting between the different wedges pushes them back,
 while expanding the mutual wedge distance is inhibited by  the osmotic pressure acting on 
the nearly touching wedges by the surrounding bacteria. The latter effect  is similar to the strong
equilibrium depletion interaction between parallel rods suspended in passive spheres~\cite{depletion,depletionNi}.
In our simulations, we observed that once the wedges are sticking together at these small distances
they are irreversibly bound during the time scale of the simulations such that we conclude that this is 
the final state of the system. Accordingly, this first peak will grow if the data are averaged over a 
longer simulation time when started from a randomly chosen distance.

\begin{figure}
\centering
\resizebox{1.0\columnwidth}{!}{\includegraphics{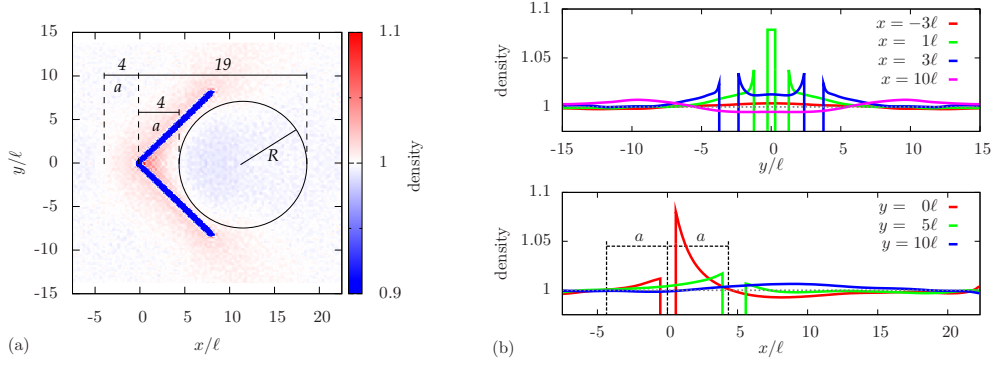}}
\caption{(a) Density profile around a single carrier indicating the depleted wake zone. Swimmer accumulation in the cusp show 
possible swirl configuration. (b) Select density profiles along {\bf x} and {\bf y} direction through the  system, showing the 
thickness $a$ of the bacteria accumulation layer.}
\label{DensityProfileSingle}   
\end{figure}

The occurrence of the next three peaks  in $P(d/\ell)$  is more subtle. 
In order to obtain a simple geometric picture for the second peak at $d_2 = 6\ell$, 
we consider the density distribution of the bacteria around a single wedge
in the frame of the moving wedge which is plotted in Fig.~\ref{DensityProfileSingle}(a). This density field reveals an accumulation
layer of thickness of about $a=4\ell$ around the wedge, see Fig.~\ref{DensityProfileSingle}(b),  and a circular depletion zone of particles 
inside the wedge~\cite{KaiserSokolov_2014}.  This depletion zone 
 possesses a typical radius $R=7.5\ell$ which coincides with the typical swirl size of the bulk bacterial suspension 
in the absence of any wedge~\cite{KaiserIEEE}. In fact, the basic idea in understanding the depletion zone is that a typical swirl swipes out particles from the inner wedge~\cite{KaiserSokolov_2014}.

\begin{figure}
\centering
\resizebox{1.0\columnwidth}{!}{\includegraphics{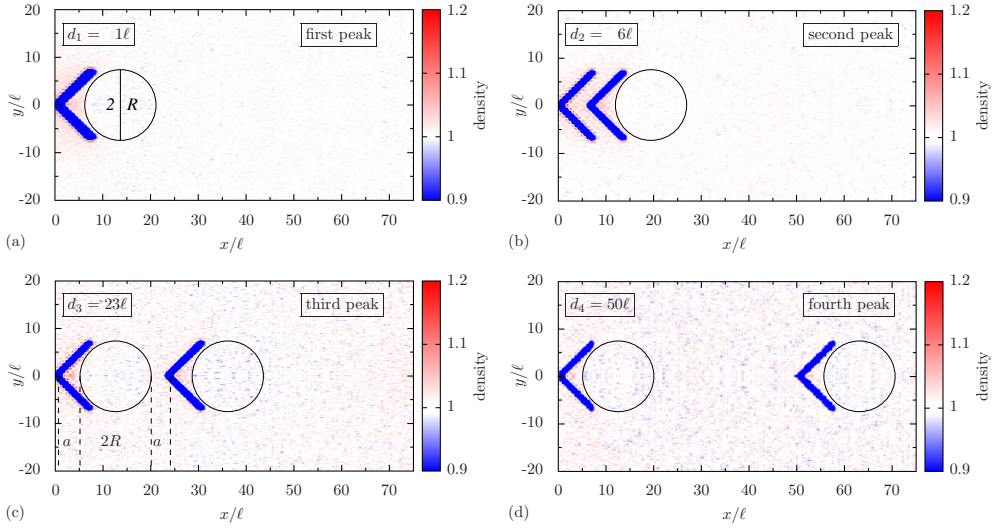}}
\caption{Density profiles around a two carriers for  four different  distances $d$ between the carriers, revealing a peak in 
the distance probability distribution $P(d/\ell)$. Circles indicate the spatial extensions of the swirls in the wake of the carriers.}
\label{DensityProfileMany}   
\end{figure} 

Figure~\ref{DensityProfileMany}  displays  the density field around two carriers 
under the constraint that they are at the distances where the four peaks in $P(d/\ell)$ occur.
Figure~\ref{DensityProfileMany}(a) corresponds to the first peak at $d_1 = 1\ell$ where the two sticking wedges can hardly be distinguished 
and the overlapping surrounding accumulation layer responsible for the depletion attraction is clearly visible.
Figure~\ref{DensityProfileMany}(b) shows the distance $d_2 = 6\ell$ where the second peak and a speed minimum occurs. 
Here two stacked wedges can make use of the depletion zone causing strong
 interpenetration. This mutual attraction is reduced when the mutual surrounding accumulation layers of bacteria
around the wedges start to overlap. This occurs roughly at a distance of $2a=8 \ell$ which is close 
to the position of the second peak at $d_2=6\ell$.

At a distance $d_3 = 23\ell$ the density field is shown in Fig.~\ref{DensityProfileMany}(c). Geometrically, as
also visualized in Fig.~\ref{DensityProfileMany}(c), this distance matches a swirl diameter
augmented by a doubled layer size $2R+2a=23\ell$ and represents the unstable situation where the
accumulation layer of the right wedge just starts to touch the inner swirl in the left wedge.
Finally, the fourth peak at $d_4=50\ell$ correlates with the occurrence of several swirls. 
However, in this distance regime, the variation in the relative wedge speed and the amplitude of the fourth peak are negligible.

\begin{figure}
\centering
\resizebox{1.0\columnwidth}{!}{\includegraphics{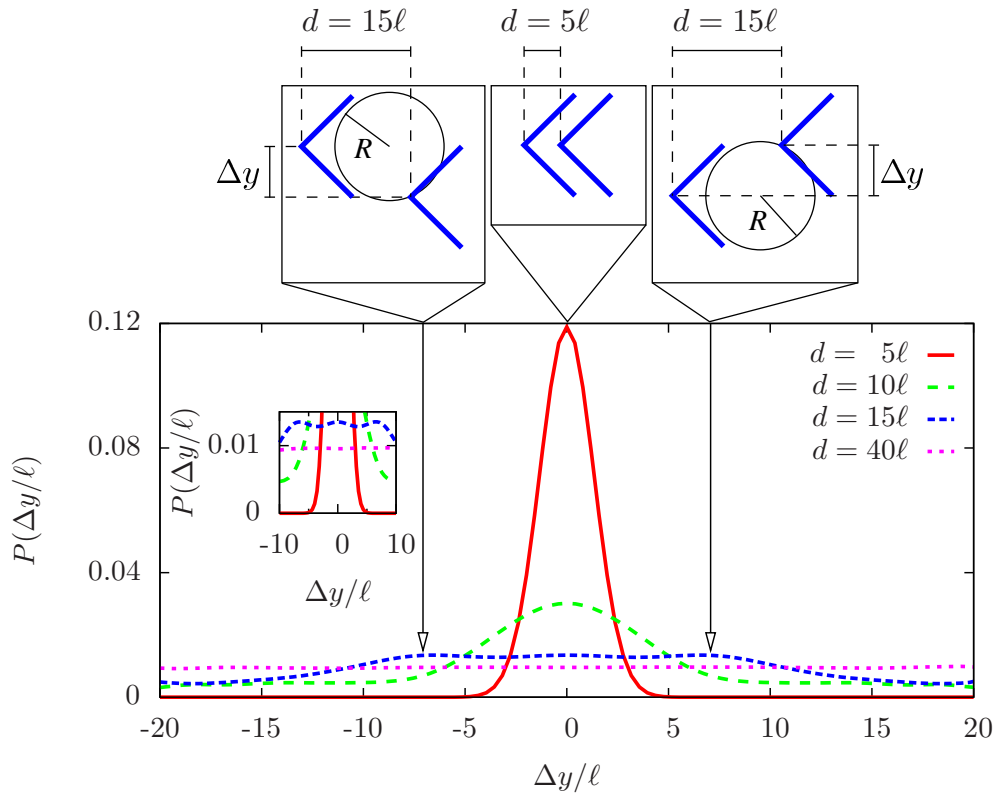}}
\caption{Probability distribution $P(\Delta y)$ for select distances $d$.  Inset shows a close up revealing the triple peak for $d=15\ell$ 
and sketches indicate the configuration for the carriers for marked distances $\Delta y$.}
\label{py}   
\end{figure} 

We  now analyze the second case relaxing the constraint in the $y$-direction.
The probability distribution $P(\Delta y /\ell)$ to find the second carrier in the perpendicular direction $\Delta y$
is shown in  Fig. \ref{py}. Obviously, this distribution is even in  $\Delta y$ due to the reflection symmetry.
For large inter-carrier distances $d$ the probability is almost uniformly distributed implying that  
motion of  two wedges is basically uncorrelated, see the data for $d=40\ell$ in  Fig. \ref{py}. As $d$ shrinks,
a triple-peaked distribution  $P(\Delta y /\ell)$ emerges, see the data for $d=15\ell$ in  Fig. \ref{py} for which the 
positions of the two wedges are also explicitly indicated. The two side peaks indicate an optimal motion
where the apex of the right wedge just experiences the outer range of the depletion zone, see Fig. \ref{DensityProfileSingle}(b).
Finally, at closer distance the motion of the right wedge is confined within the aperture of the left one 
resulting in a localized  distribution function $P(\Delta y /\ell)$. Again, the observed fine structure supports the 
general swirl depletion picture put forward in Ref.~\cite{KaiserSokolov_2014}.

\section{Experiment}
\label{resultsExp}

\begin{figure}
\centering
\resizebox{1.0\columnwidth}{!}{\includegraphics{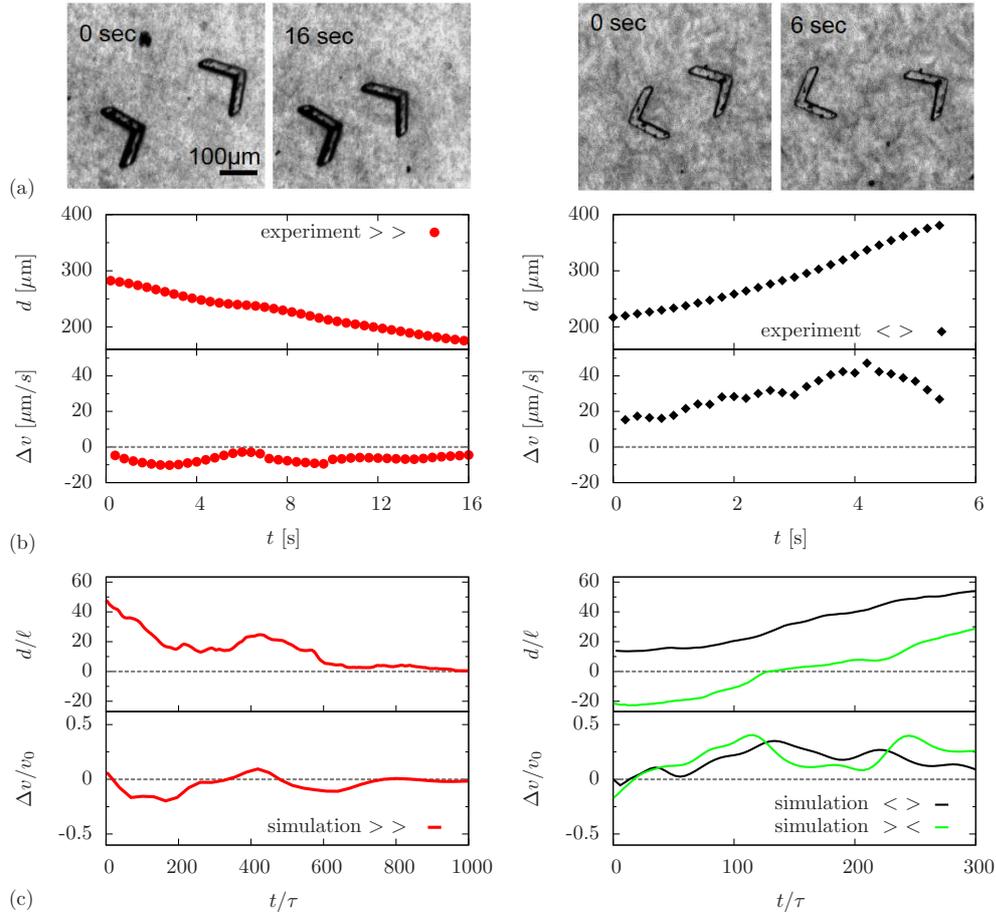}}
\caption{(a) Illustration of the attraction and repulsion of wedge-like carriers in a turbulent bacterial bath.
The orientation of the wedges is controlled by the applied uniform magnetic field.
Distance $d$ between carriers and the respective relative velocity $\Delta v$ for parallel $>\ >$ (red) and anti-parallel orientation $<\ >$ 
(black) as a function of time obtained from (b) experiments and (c) simulations. 
A negative sign in $d$ indicates a situation of opposed wedges~$> \ < $.}
\label{ExpFig}   
\end{figure} 

We perform series of experiments to support our theoretical findings. The micro-wedges were fabricated by photolithography from a
mixture of photoresist and nickel particles. Nickel particles conglomerate into long chains in the course of  spin coating of the 
mixture on a silicon wafer. After the exposure, developing, and etching, the wedgelike carriers containing small nickels particles 
were transferred  to water by ultrasonication. Due to magnetization of the particles, the orientation of the fabricated wedges can 
be controlled by the external magnetic field. Two orthogonal pairs of large Helmholtz coils were used to control the orientation of 
the carriers. The uniform field created by these coils does not affect the positions or horizontal motions of the wedges, but only 
control their orientation.  
The micro-wedges were carefully placed into a bacterial suspension by pipetting. For this purpose the bacteria {\it Bacillus subtilis}
were picked from a single colony on an agar plate and placed in a plastic tube filled with Terrific Broth growth medium. For optimal growth
the bacteria were incubated at $35^{\circ}$C for 12 hours. Before the experiments the bacteria were extracted from the growth medium, washed
and concentrated by centrifugation. The experiments were performed in a free-standing liquid film of 200-400 micron thickness, see details 
in Ref.~\cite{2007SoEtAl}.

The dynamics of the wedges in the bacterial suspension was captured by Olympus IX71 microscope and digital 
monochrome camera (Procilica GX 1660), see Fig.~\ref{ExpFig}(a). 
In Figs.~\ref{ExpFig}(b),(c) we compare experimental and numerical results for the temporal progress of the distance between two
carriers for different configurations as well as the resulting velocity difference $\Delta v = \Delta d / \Delta t$. We observed 
convergence of carriers of the same orientation and repulsion in the case of the 
opposite orientations, see Fig.~\ref{ExpFig}(b), which supports our simulation results, see Fig.~\ref{ExpFig}(c). 
At distances less than 10 micron (i.e. of the order of two bacteria length) the  
magnetic interaction between the shuttles dominates over hydrodynamic interaction. If the particles collide, they usually stick 
to each other at random orientations. 

To mimic a collision of two carriers opposed in orientation ($> \ < $) we have performed further simulations 
for an initial anti-parallel configuration, see Fig.~\ref{ExpFig}(c).
Since each wedge is transported in the direction along its cusp, they will approach each other, collide, slide along each other and finally drift apart. 
Concomitantly, we observe a slow-down in the relative velocity during the collision process.
We also emphasize another  difference between experiment and numerical simulations. While the motion of carriers is confined in two 
dimensions, the motion of bacteria is three-dimensional. As a result, we also observed large (compared to simulations) fluctuations 
in the positions of carriers due to bacterial activity. These fluctuations often prevent sticking of the carriers. 

\section{Conclusions}
\label{conclusions}
In line with the fascinating topic of how many passive objects self-organize
in an active fluid, we have considered here the case of two micro-wedges with fixed orientations
in a turbulent bacterial bath. We find an efficient stacking of the two wedges of same orientation
leaving no bacteria between them. This state is a stable dynamical mode where the
two micro-wedges are following each other with the same speed. There is more subtle
behavior in the relative wedge motion which is compatible with the geometric swirl depletion picture 
put forward in Ref.~\cite{KaiserSokolov_2014}. Our findings provide a first step
towards the general case of many carriers which are therefore expected to form
columnar stacks with a  large persistence length reminiscent to the columnar phase of
stacked bowl-shaped colloidal particles~\cite{Marechal0}.

For the future study, it would be interesting to investigate 
the influence of hydrodynamic interactions~\cite{GoldsteinPNAS2014} and the dynamics of submersed passive particles 
whose motion is non-restricted, as well as other
particle shapes such as $L$, $C$ shapes~\cite{BtH_L_part,Wensink2014} where stacking is also
expected. However, there are also shapes where stacking is frustrated
(like for $T$-shaped carriers) which are expected to form loosely-packed gels~\cite{Dijkstra}.
Furthermore, for future research, it would be challenging to study the motion of passive particles for 
gliding bacteria, where large clusters emerge~\cite{PerunaiPRL12}. It is expected that such clusters have a significant 
influence on the dynamics of the wedges. For gliding bacteria,
hydrodynamic interactions are less relevant which makes our modelling an even more appropriate one for this realization. 
Moreover, a microscopic theory for active depletion~\cite{Egorov} is highly
desirable to make predictions
for the carrier motions which could be based on kinetic~\cite{Ihle}
or dynamical density functional theory~\cite{Wensink2008,WittkowskiMolPhys2011}.
%

\begin{acknowledgement}
A.~K. was supported by the ERC Advanced Grant INTERCOCOS (Grant No. 267499) and
H.~L. by the science priority program SPP 1726 of the German Science Foundation (DFG). 
Work by A.~S. and I.~S.~A. was supported by the U.S. Department of Energy (DOE), Office of Science,
Basic Energy Sciences (BES), Materials Science and Engineering Division.
\end{acknowledgement}

\bibliography{./refs}
\end{document}